\documentclass[final,5p,times,twocolumn]{elsarticle}
\usepackage{graphicx}
\usepackage{caption}
\usepackage{latexsym}
\usepackage{amsmath,amssymb}
\usepackage{lipsum}
\usepackage{xcolor}
\usepackage[utf8]{inputenc}
\usepackage[english]{babel}
\usepackage{lineno}
\usepackage{comment}
\usepackage[T1]{fontenc}
\usepackage{tikz}
\usepackage{commath}
\usepackage[colorlinks=True,linkcolor=DarkRed,citecolor=ForestGreen,urlcolor=MediumBlue,
]{hyperref}
\graphicspath{{figures/}}

\def \suppsec[#1]{SI Section}
\def \SF[#1]{Suppl.~Fig.~#1}
\def \F[#1]{Fig.~#1}

\def \todo[#1]{\color{red}#1\color{black}}

\usepackage{amsmath}

\addto\captionsenglish{}
\usepackage{bibunits} 
\usepackage{titlesec}
\titleformat{\paragraph}[runin]
{\bfseries\scshape}{\theparagraph}{1em}{}

\titleformat{\section} {\bfseries}{\thesection}{1em}{}
\titleformat{\subsection} {\bfseries}{\thesubsection}{1em}{}

\makeatletter
\def\ps@pprintTitle{} 
\makeatother

\biboptions{sort&compress} 

\begin{document}
\setlength\linenumbersep{0.1cm}
\newcommand{\markerone}{\raisebox{0.5pt}{\tikz{\node[draw,scale=0.4,circle,fill=black!0!blue](){};}}}
\newcommand{\markertwo}{\raisebox{0.5pt}{\tikz{\node[draw,scale=0.4,circle,fill=black!0!white](){};}}}

\title{Vortex core spectroscopy links pseudogap and Lifshitz critical point in a cuprate superconductor}

\author[1]{Tejas Parasram Singar}
\author[1]{Ivan Maggio-Aprile}
\author[2]{Genda Gu}
\author[1]{Christoph Renner}
\address[1]{Department of Quantum Matter Physics, University of Geneva, 1211 Geneva, Switzerland}
\address[2]{Condensed Matter Physics and Materials Science Department,
Brookhaven National Laboratory, Upton, New York 11973, USA}

\date{\today}

\begin{abstract}
Understanding how superconductivity competes with other electronic phases in cuprates requires direct access to the hidden non-superconducting low temperature phase, for which Abrikosov vortices provide a unique local probe. We map the doping- and field-dependent evolution of vortex-core states in Bi$_{2}$Sr$_{2}$CaCu$_{2}$O$_{8+\delta}$ across a broad doping range spanning the Fermi-surface Lifshitz transition. High-resolution scanning tunneling spectroscopy reveals a striking transformation of the vortex-core spectrum from unconventional, pseudogap-like signatures at moderate doping to more BCS-like behavior beyond a critical doping $p^* \approx 0.21$. This crossover aligns with the pseudogap endpoint and the onset of Fermi-surface reconstruction, indicating a direct link between pseudogap physics and vortex electronic structure. Our findings highlight the vortex core as a sensitive local probe of the cuprate ground state.
\end{abstract}

\maketitle

\section{Introduction}
The mechanism responsible for superconductivity at unprecedented high-temperatures discovered nearly forty years ago in a copper oxide~\cite{Bednorz1986}, remains largely unexplained. Although many characteristic features of high temperature superconductors (HTS) have been studied extensively~\cite{fischer2007}, the origin of electron pairing is yet to be explained. Among the outstanding questions likely to contribute to understanding the pairing mechanism is the electronic structure of Abrikosov vortices, bundles of quantized magnetic flux penetrating a type II superconductor when it is immersed in an external magnetic field. As we show in the present study, the vortex core, where superconductivity is locally suppressed by the magnetic field, provides a unique window into the underlying pseudogap and normal state that is otherwise hidden at low temperatures.

Abrikosov vortices are essential for practical applications and for advancing our fundamental understanding of superconductivity. They play a critical role in superconducting magnets, magnetic levitation, and superconducting power cables, where optimizing vortex pinning is crucial to minimizing energy dissipation and enhancing performance. From a fundamental perspective, they serve as a microscopic probe of the normal and superconducting ground states, revealing details about the pairing symmetry, competing phases, quasiparticle dynamics, and phase coherence.

A vortex consists of a core region, where superconductivity is suppressed over a distance of the order of the superconducting coherence length $\xi$ from its center. Each vortex carries one quantum of magnetic flux that is screened by superfluid currents circulating and decaying around the core over a distance known as the London penetration depth $\lambda$.  

\begin{figure*}[htb]
    \centering
   \includegraphics[width=2\columnwidth] {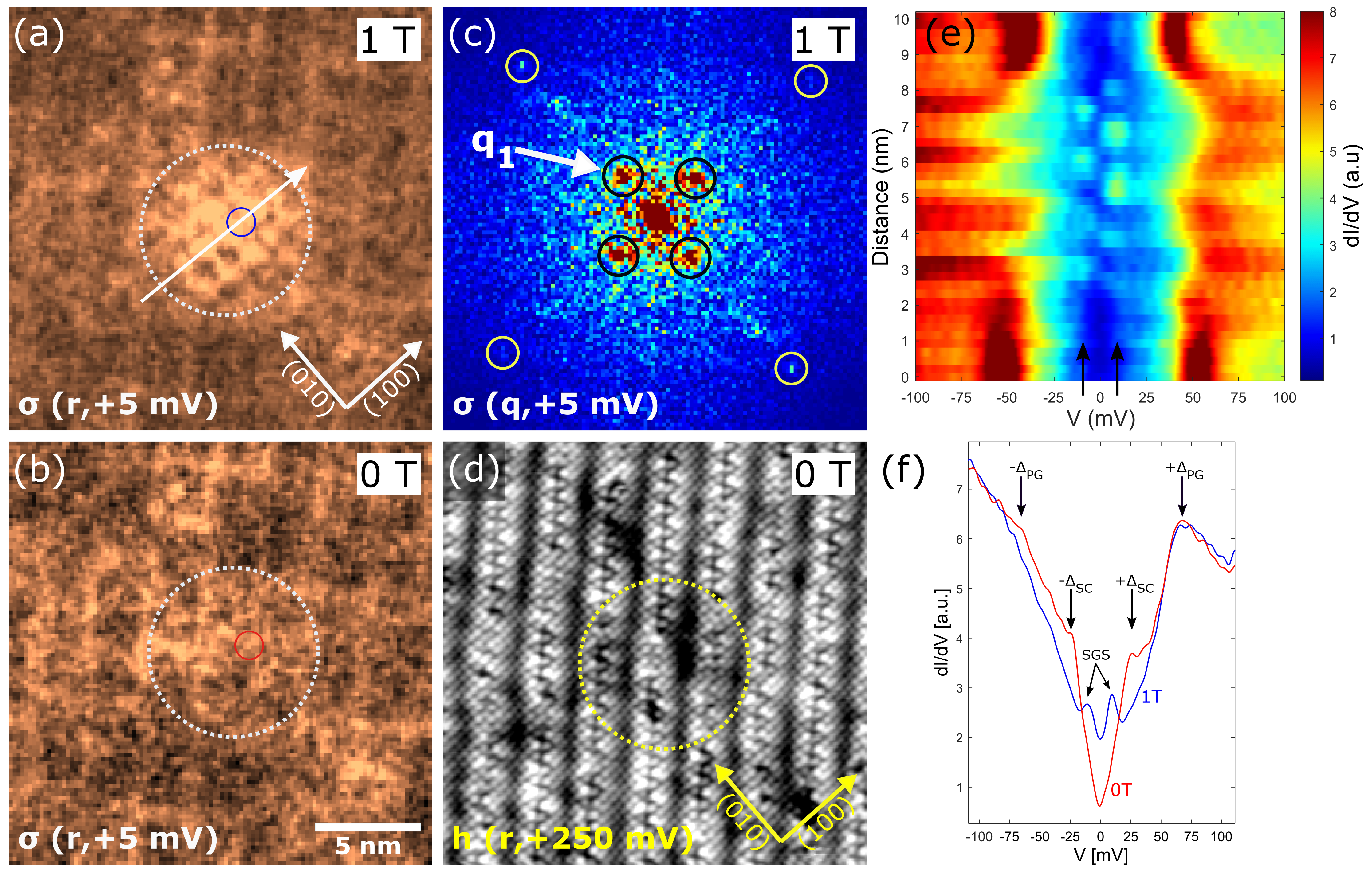}
    \caption{\textbf{Electronic vortex core structure in Bi-2212 at $p < p^*=$~0.21.} $dI/dV(r, 5 mV)$ conductance maps acquired in the exact same field of view at (a) 1~Tesla and (b) 0~Tesla. (c) Fourier transform of panel (a) with yellow and black circles indicating the atomic Bragg peaks and the $q_{1}$ checkerboard peaks, respectively. (d) Topography acquired in the same field of view as panels (a) and (b) at 0~T ($V_{\text{b}} =$ 250~mV, $I =$ 20~pA, dashed circle identifies the vortex position). (e) Tunneling spectra as a function of position along the arrow in panel (a) showing the spatial dependence of the SGSs. (f) $dI/dV(V)$ spectra averaged over the regions marked by a solid blue circle at 1~Tesla in (a) and a solid red circle at 0~Tesla in (b). The corresponding local doping is $p_{\text{local}}= 0.13\pm0.02$; the bulk $T_{c}\approx$ 80~K; Data was acquired at 4.4 K, with $V_{\text{b}} =$ 110~mV and $I =$ 50~pA.}
    \label{fig: CBVTXUD80K}
\end{figure*}

The electronic vortex core structure of conventional low temperature superconductors has been described in details by Caroli, de Gennes, and Matricon (CdGM)~\cite{caroli1964}. Their prediction of quasiparticle bound states at discrete energy levels for a \textit{s}-wave symmetric superconducting gap ($\Delta$) has been confirmed experimentally by scanning tunneling spectroscopy (STS) in a number of conventional superconductors. First by Hess et al.~\cite{hess1989} in 2H-NbSe$_2$, who emphasized band structure effects and competition with a charge density wave (CDW)~\cite{hess1990}. Many compounds with a larger $\Delta^2/E_F$ ($E_F$: Fermi level) have since been investigated, allowing exquisite energy and spatial resolution of the quasiparticle bound states structure in agreement with the CdGM predictions for \textit{s}-wave superconductors~\cite{suderow2014, chen2018, chen2020}. 

The gap function in Bi$_{2}$Sr$_{2}$CaCu$_{2}$O$_{8+\delta}$ (Bi-2212) has \textit{d}-wave symmetry, with a maximum gap amplitude along the Cu-O bonds. The corresponding electronic vortex core structure is expected to be very different due to the presence of nodes in the gap, with a continuum of states rather than discrete bound states~\cite{wang1995}. Model calculations predict a conductance peak near the Fermi level instead of a gap of the order of $\Delta^2/E_F$ expected in the \textit{s}-wave case~\cite{caroli1964}. However, early scanning tunneling spectroscopy of the vortex cores failed to observe such a \textit{d}-wave signature~\cite{renner1998, hoogenboom2000, pan2000, hoffman2002, levy2005, matsuba2007, yoshizawa2013, machida2016}. STS found instead characteristic subgap states (SGSs) and a $\approx 4a_{o} \times 4a_{o}$ checkerboard charge modulation in the vortex halo (where $a_{o}$ is the crystallographic unit cell). SGSs have also been observed in YBa$_{2}$Cu$_{3}$O$_{7-\delta}$~\cite{maggio1995} and HgBa$_{2}$Ca$_{2}$Cu$_{3}$O$_{8+\delta}$~\cite{ZhangPRB2024}, and vortices in Ca$_{2-x}$Na$_{x}$CuO$_{2}$Cl$_{2}$~\cite{hanaguri2009} show SGSs and a checkerboard, suggesting some universality of these features. 

The zero-bias conductance peak (ZBCP) expected for \textit{d}-wave superconductors was only recently uncovered in YBa$_{2}$Cu$_{3}$O$_{7-\delta}$~\cite{bruer2016} and measured directly in heavily overdoped Bi-2212~\cite{gazdic2021,MAGGIOAPRILE2023}. Here, we present a systematic characterization of the vortex core structure in Bi-2212 as a function of hole doping. We find two classes of vortex cores depending on hole doping, with subgap states and periodic charge modulations characteristic of vortices located in regions with a doping $p<0.21$ holes per unit cell, and a ZBCP and no local charge modulations in vortices located in regions with $p>0.21$, regardless of the magnetic field strength.

\section{Electronic vortex core structure}
Scanning tunneling spectroscopy is sensitive to the local density of states (LDOS), which is modified over a coherence length inside the vortex core. This enables to image vortices by mapping the LDOS as a function of position \cite{MAGGIOAPRILE2023}. The vortex cores discussed here were measured at 4.4~K in a magnetic field applied perpendicular to the CuO planes. We selected Bi-2212 for this study because of the ease with which atomically clean surfaces can be prepared for local probe experiments by in situ cleaving (fig.~\ref{fig: topovariation}), and because of the knowledge accumulated over four decades of research on this compound. Ideally, the doping dependence of the vortex core structure would be studied using tunable space charge doping of exfoliated thin crystals~\cite{SterpettiIOP2019}. However, the preparation of suitable devices with surfaces clean enough for STS has proven difficult. We therefore rely on the availability of single crystals covering a wide range of hole doping. The results discussed here were obtained on four batches of Bi-2212 single crystals: two underdoped with $T_c=52$~K (UD52) and $T_c=80$~K (UD80), and two overdoped with $T_c=70$~K (OD70) and $T_c=52$~K (OD52).

Because the doping is inhomogeneous in Bi-2212~\cite{mcelroy2005}, defining the nominal doping based on $T_c$ is not suitable. A more accurate estimate of the hole concentration $p$ at the position of each vortex core is required. To this end, we use the local gap amplitude in the vicinity of each vortex and equation $E_{\text{pg}}=E_{\text{pg}}^{\text{max}}(0.27 - p)/0.22$, where $E_{\text{pg}} = 2\Delta_{\text{pg}}$ and $E_{\text{pg}}^{max} = 152\pm 8$~meV \cite{Hufner2008}. In our experiments, $\Delta_{\text{pg}}$ is the pseudogap amplitude extracted from the average of the tunneling spectra measured within a radius of 5.4~nm around the vortex center but outside the core region. Above $p=0.21$ where the pseudogap merges into the superconducting gap in the tunneling spectra~\cite{Hufner2008}, we use the same equation to determine $p$, but replacing $\Delta_{\text{pg}}$ with the superconducting gap $\Delta_{\text{sc}}$ extracted from the tunneling spectra. 

\begin{figure}[t]
    \centering
    \includegraphics[width=\columnwidth] {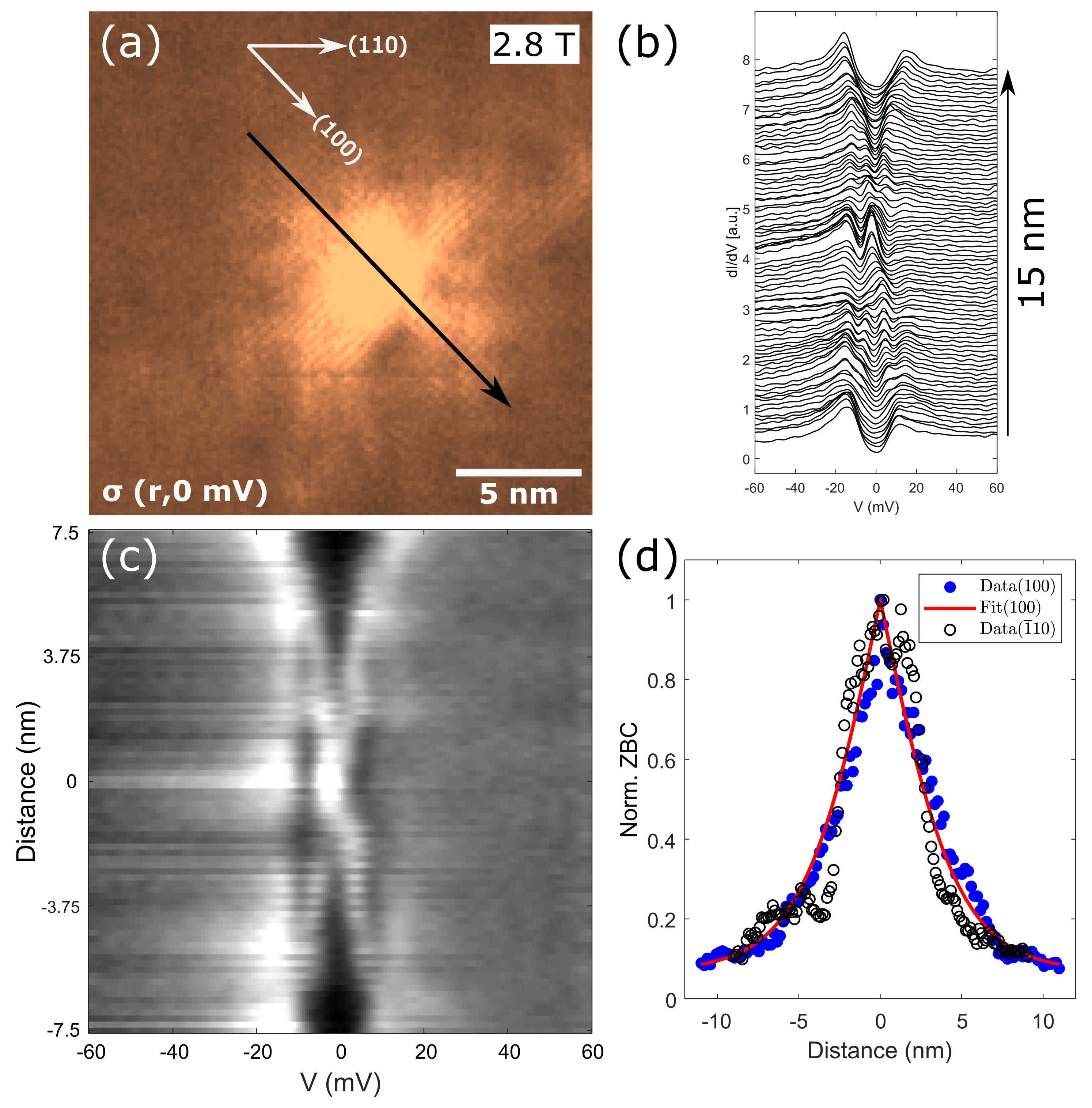}
    \caption{\textbf{Electronic vortex core structure in Bi-2212 at $p > p^*=$~0.21.} (a) $dI/dV(r, 0~mV)$ conductance map measured at 2.8~T. (b) Tunneling spectra as a function of position and (c) corresponding gray-scale plot measured along the black arrow in panel (a). (d) Zero-bias conductance as a function of distance from the vortex center along the (100) direction ($\textcolor{blue}{\bullet}$) and along the ($\overline{1}$10) direction ($\circ$). The red curve is a fit to $\sigma(x) = 1 - (1 - \sigma_0) \cdot \tanh(x/\xi)$ of the data along the (100) direction, where $\sigma_0$ is the normalized zero bias conductance away from the vortex region and $x$ is the distance from the core center~\cite{EskildsenPRL2002}. The local doping is $p_{\text{local}}= 0.24\pm0.01$; Data was acquired at 4.4~K, with $V_{\text{b}} =$ 110~mV and $I =$ 50~pA.}
    \label{fig: OD52KZBCPdecay}
\end{figure}

To characterize the doping dependence of the vortex core in a \textit{d}-wave superconductor, we start with STS data measured on UD80 in a magnetic field of 1~Tesla. These experiments reveal all the characteristic features reported in the literature for a vortex in underdoped Bi-2212 (Fig.~\ref{fig: CBVTXUD80K}). In the tunneling conductance map at 5~mV (Fig.~\ref{fig: CBVTXUD80K}a), we find an approximately $4a_{o} \times 4a_{o}$ charge modulation spanning the vortex halo, with corresponding $q_1$ peaks resolved in its Fourier transform (Fig.~\ref{fig: CBVTXUD80K}c). The low resolution of the $q_1$ peaks is a consequence of the small area of the vortex halo where the periodic modulation is present. The latter is associated with SGSs alternating between negative and positive bias as a function of position through the vortex core (Fig.~\ref{fig: CBVTXUD80K}e), consistent with earlier reports (see e.g. \cite{levy2005,matsuba2007}). Because of this spatial bouncing, the SGSs will appear as symmetric or asymmetric peaks in individual tunneling spectra, depending on spatial averaging and precise position of the tip with respect to the vortex core. 

Repeating the same conductance map on exactly the same region at zero-field reveals the same background conductance as in Fig.~\ref{fig: CBVTXUD80K}a, with faint LDOS stripes oriented along the (110) superstructure direction (Fig.~\ref{fig: CBVTXUD80K}b) but no $\approx 4a_{o} \times 4a_{o}$ periodic modulation at low energy (fig.~\ref{fig: UD80KVtx4suppl}). Figures~\ref{fig: CBVTXUD80K}a and \ref{fig: CBVTXUD80K}b were aligned with atomic precision based on high-resolution topographic images of the same field of view (Fig.~\ref{fig: CBVTXUD80K}d). The presence of a vortex is identified by the vanishing of the superconducting coherence peaks $\pm \Delta_\text{SC}$ and the emergence of SGSs, two characteristic changes in the tunneling conductance spectra in presence of a vortex core (Fig.~\ref{fig: CBVTXUD80K}f). On the other hand, the pseudogap features at $\pm \Delta_\text{PG}$ do not change significantly at the center of a vortex.

Turning to OD52 crystals in a magnetic field of 2.8~Tesla, we find a vortex core structure (Fig.~\ref{fig: OD52KZBCPdecay}) that is remarkably different from the underdoped case examined in Fig.~\ref{fig: CBVTXUD80K}. Most striking differences are the absence of a periodic charge modulation in the vortex halo (Fig.~\ref{fig: OD52KZBCPdecay}a) and the presence of a clear ZBCP at the vortex center (Fig.~\ref{fig: OD52KZBCPdecay}b, fig.~\ref{fig: OD52KZBCP-vtxpush}). This peak splits into two low energy states at positive and negative bias, which shift away from $E_F$ with increasing distance from the vortex center (Fig.~\ref{fig: OD52KZBCPdecay}c). 

The electronic structure of the vortex core in Fig.~\ref{fig: OD52KZBCPdecay} is structurally and quantitatively remarkably consistent with the theoretical predictions for a \textit{d}-wave superconductor \cite{wang1995}, including a weak nodal versus antinodal anisotropy \cite{berthod2017}. The coherence length $\xi$ can be extracted by fitting the normalized zero-bias conductance $\sigma_0$ as a function of position $x$ to $\sigma_0(x) = 1 - (1 - \sigma_{\infty}) \cdot \tanh{x/\xi}$, where $\sigma_{\infty}$ is the zero bias conductance away from the vortex. We obtain a coherence length $\xi \approx 4.7$~nm in this case (Fig.~\ref{fig: OD52KZBCPdecay}d). A precise analysis shows that $\sigma_0(x)$ decays slightly faster along the nodal direction ($\overline{1}$10) than along the anti-nodal direction (100). 

\section{Discussion}
\begin{figure}
    \centering
    \includegraphics[width=\columnwidth] {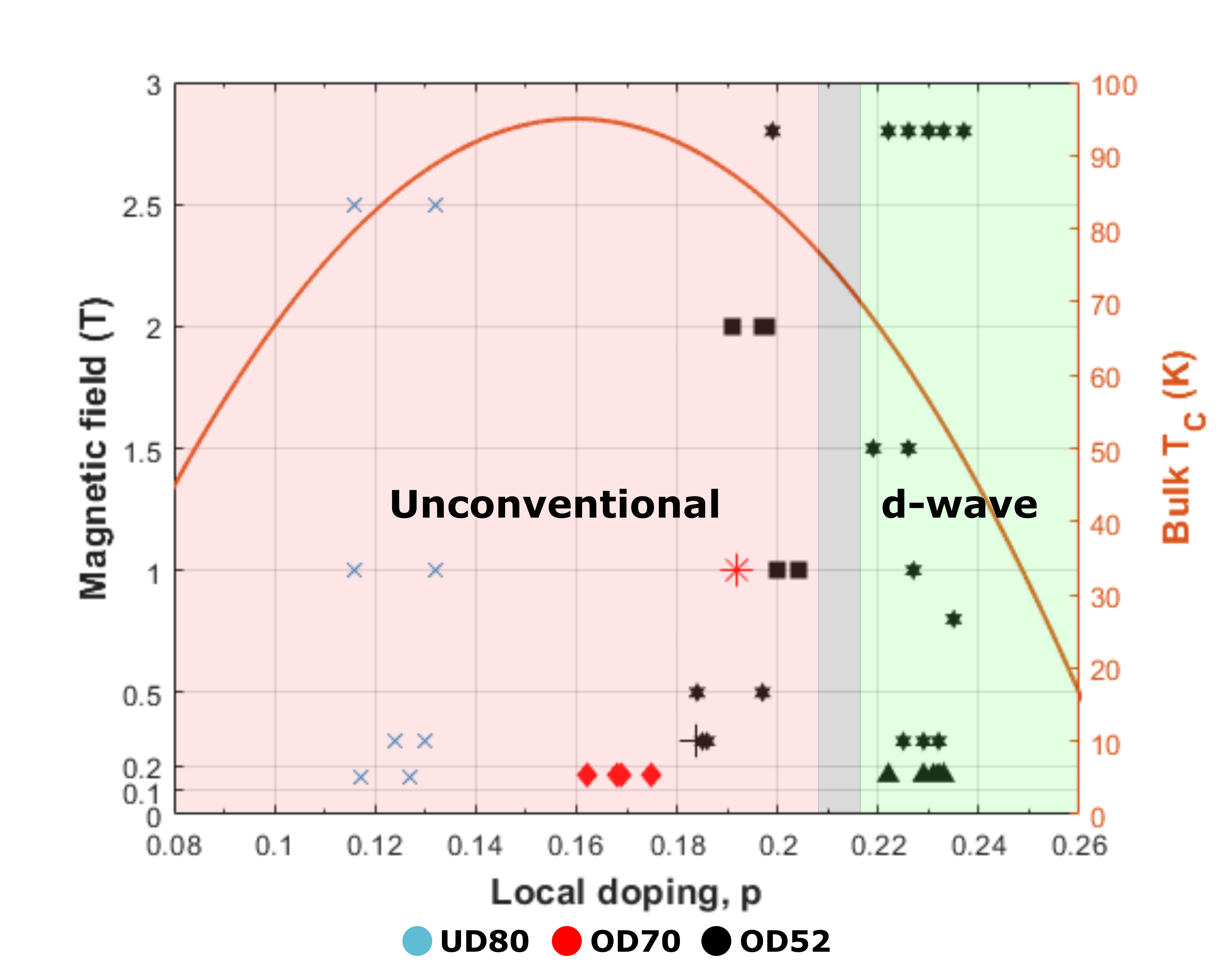}
    \caption{\textbf{Field-Doping phase diagram of Abrikosov vortex cores in hole-doped Bi-2212.} Each data point corresponds to a complete $dI/dV(r, V, B)$ grid map with a vortex in the field of view. Different marker colors represent different bulk $T_{\text{c}}$ values, whereas individual samples are distinguished by distinct symbols. The local doping is extracted from the gap distribution outside the vortex core region within a radius of $2\xi$ where $\xi \approx$ 2.7~nm is the coherence length. The bulk $T_{\text{c}}$ curve corresponds to $T_{\text{c}} = T_{\text{c}}^{\text{max}}[1 - 82.6(p-0.16)^{2}]$ where $T_{\text{c}}^{\text{max}} =$ 95~K at $p = p_{\text{opt}} =$ 0.16, following Presland \textit{et al.} \cite{PRESLAND1991}.}
    \label{fig: phasediagram}
\end{figure}

We have examined over forty high-resolution STS conductance maps on UD80, OD70 and OD52 single crystals, each containing at least one vortex (fig.~\ref{fig: Doping_uncertainty}). Using this large dataset measured in magnetic fields between 0.15 and 2.8~Tesla at 4.4~K, we construct\textit{} the low-temperature vortex $H-p$ phase diagram in Fig.~\ref{fig: phasediagram}, over a wide range of hole concentrations from $p=$ 0.12 to $p=$ 0.24. We were unfortunately not able to identify vortices at lower doping in UD52 due in part to the presence of strong $\approx 4a_{o} \times 4a_{o}$ conductance modulations near $E_F$ even without any applied magnetic field (fig.~\ref{fig: UD52KCB}). 

The phase diagram in Fig.~\ref{fig: phasediagram} graphically illustrates the key result of this study: all vortex cores belong to one of two classes. Vortices of the first class, which occur at $p<$ 0.21, exhibit unconventional properties highlighted in Fig.~\ref{fig: CBVTXUD80K}. They show $\approx 4a_{o} \times 4a_{o}$ conductance modulations with SGSs, while the zero-bias conductance peak expected for a \textit{d}-wave superconductor~\cite{wang1995} is systematically absent. Vortices of the second class, observed at $p>$ 0.21, show the electronic structure expected for a \textit{d}-wave superconductor discussed in Fig.~\ref{fig: OD52KZBCPdecay}, without SGSs or periodic conductance modulations. Note that all the vortices we have characterized belong to one of these two classes. We did not find any vortex that would indicate the simultaneous presence of both characteristics. 

The class to which a vortex core belongs does not seem to be affected by magnetic field intensities up to 2.8 Tesla, and even beyond that if we take into account data from the literature~\cite{machida2016, edkins2019, levy2005}. It is determined by doping, with the boundary separating the two classes in Fig.~\ref{fig: phasediagram} located near $p^*=0.21$ holes per unit cell. This coincides with the doping at which the pseudogap has been reported to disappear~\cite{BenhabibPRL2015, LoretPRB2017, LoretPRB2018, Doiron-LeyraudNatComm2017} and where signatures of critical quantum behavior have been found by Raman scattering~\cite{Auvray2019} and optical conductivity~\cite{park2025arxiv}. It also coincides with the Lifshitz transition, where the Bi-2212 Fermi surface evolves from open hole-like ($p$) to closed electron-like ($1+p$) \cite{KaminskiPRB2006, PutzkeNatPhy2021}. However, we detect no changes in the nature of the vortex core structure at $p=0.16$, corresponding to optimal doping, or at $p=0.19$, identified as a quantum critical point (QCP) by resonant inelastic x-ray spectroscopy~\cite{ArpaiaNatComm2023}, angle resolved photoemission spectroscopy~\cite{ChenScience2019, VishikPNAS2012} and specific heat experiments~\cite{Tallon2022}.

Note that while we observe only checkerboard vortices in nominally underdoped (UD80) and moderately overdoped (OD70) Bi-2212, both types of vortices are found at different locations in heavily overdoped crystals (OD52) (Fig.~\ref{fig: OD52KZBCPdecay}, fig.~\ref{fig: CBOD52vortex}). This is a direct consequence of the inhomogeneous hole doping, and implies that relying solely on the bulk crystal critical temperature $T_c$ to determine $p$ can be misleading and would certainly not allow to determine the phase diagram in Fig.~\ref{fig: phasediagram}. Crucially, this means that bulk techniques may not reliably identify critical doping levels such as $p^*$, since they likely average over a range of doping levels considering the doping distribution visualized in fig.~\ref{fig: sqrt2modulation}.

\begin{figure}
    \centering
    \includegraphics[width=\columnwidth] {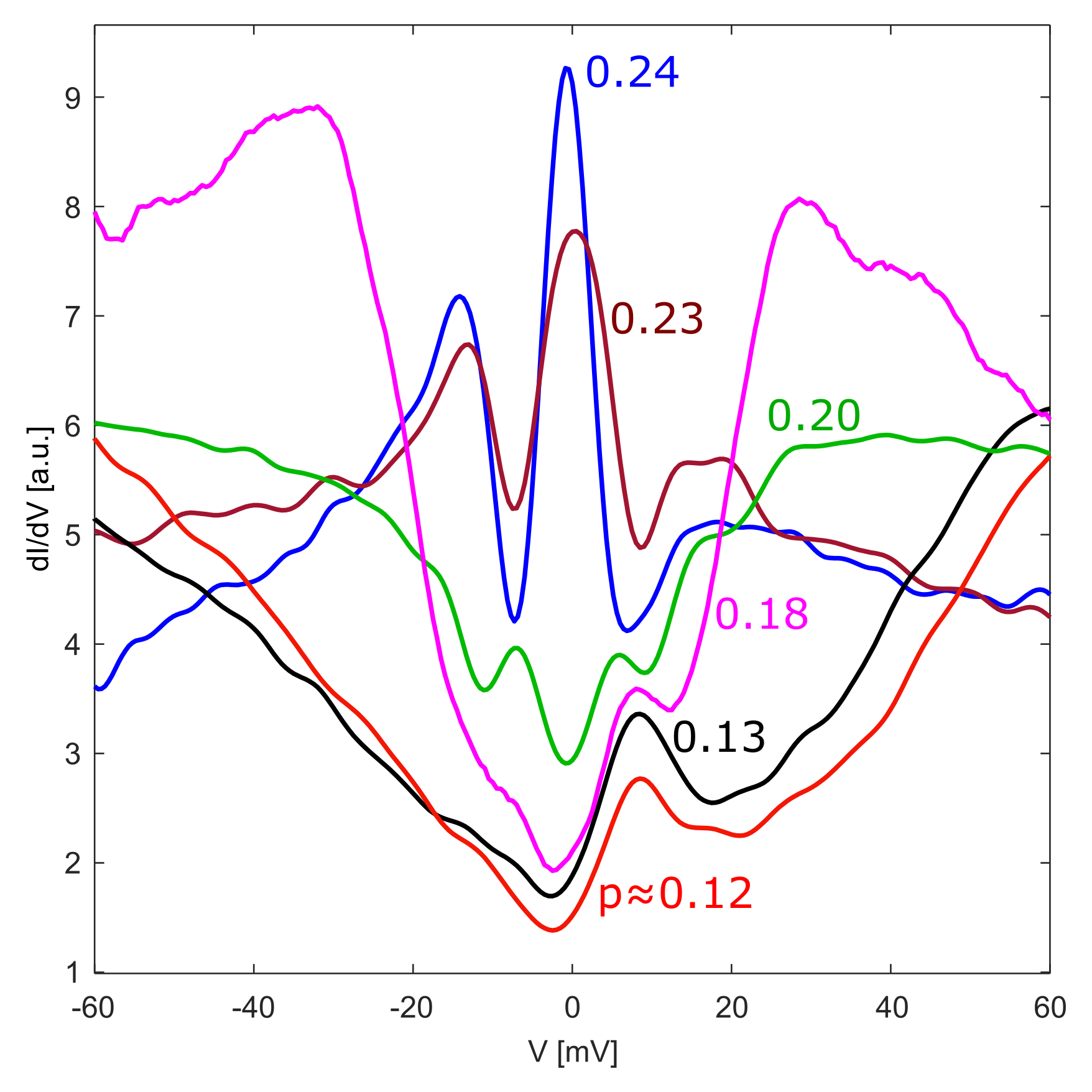}
    \caption{\textbf{Tunneling conductance spectra measured at the center of vortex cores in Bi-2212 as a function of local doping $p$.} Spectra are spatially averaged, normalized, and vertically offset for clarity.}
    \label{fig: corests}
\end{figure}

The changing vortex core structure in the vicinity of $p^*$ has been addressed theoretically in terms of strong electronic correlations \cite{DattaPRB2023, liuPRR2023} and vanishing pseudogap \cite{zhangSachdevPRB2024}. These calculations are in qualitative agreement with the changing tunneling conductance as a function of doping shown in Fig.~\ref{fig: corests}, where SGSs and pseudogap features at $p<p^*$ give way to a ZBCP at the center of the cores in regions where $p>p^*$. Our tunneling experiments clearly show that the pseudogap phase extends to $p^*\approx 0.21$, where the vortex cores develop conventional \textit{d}-wave characteristics. We further observe a modification of the periodic charge modulations across $p^*$ in the absence of magnetic field, consistent with a Lifshitz transition~\cite{KaminskiPRB2006}. They change from a $\approx 4a_{o} \times 4a_{o}$ pattern, corresponding to $q_1$ scattering along ($\pi,0$)  on the hole-like Fermi surface, to a $\sqrt{2}a_{o} \times \sqrt{2}a_{o}$ pattern (fig.~\ref{fig: sqrt2modulation}) favored by ($\pi,\pi$) scattering on the electron-like Fermi surface~\cite{ZouNatPhy2022}. This change is concomitant with the disappearance of the periodic charge modulations observed in the vortex halo as the ZBCP develops above $p^*$. 

The vortex core structure found in highly overdoped Bi-2212 is fully consistent with theoretical predictions for a \textit{d}-wave superconductor~\cite{wang1995}. However, the microscopic origin of the low-energy vortex core checkerboard pattern found in regions where $p< 0.21$ is unclear and has been a topic of intense scrutiny. Machida \textit{et al.}~\cite{machida2016} attribute it to the interference patterns caused by the scattering of Bogoliubov quasiparticles in the nodal region. However, the spatially bouncing, particle-hole asymmetric core SGSs seen in Fig.~\ref{fig: CBVTXUD80K}e have been suggested by Zhang and Sachdev \cite{zhangSachdevPRB2024} to be a consequence of a CDW order. Finally, Yu-Shiba-Rusinov excitations have been reported close to magnetic impurities~\cite{menard2015}, where spatially modulated states are strikingly similar to the ones we detect in the vortex halos. Which interpretation is correct is beyond the scope of the present study and is part of ongoing experiments.

\section{Conclusion}
The local probe vortex core spectroscopy discussed here provides unique access to the low temperature electronic properties of Bi-2212 hidden beneath the superconducting dome, where the superconducting phase cannot be easily suppressed. Our results provide direct local access to the evolution of the Fermi surface and local density of states as a function of doping. We find compelling evidence that the pseudogap in Bi-2212 persists up to a critical doping of approximately $p^*=0.21$, where the Fermi surface topology changes from open hole-like to closed electron-like. This Lifshitz transition is corroborated by a radical change in the doping dependent quasiparticle interference patterns, which change from $\approx 4a_{o} \times 4a_{o}$ to $\sqrt{2}a_{o} \times \sqrt{2}a_{o}$ across $p^*$. Concomitantly to the disappearance of the pseudogap at the Lifshitz transition, the vortex cores change to the electronic structure expected for a \textit{d}-wave superconductor at $p^*$. Our detailed doping dependent high resolution scanning tunneling spectroscopy sheds new light on the doping phase diagram of Bi-2212, and establishes the vortex core as a sensitive local probe of the cuprate ground state. It resolves ongoing discussions about the extent of the pseudogap, establishing its persistence up to the Lifshitz transition, and highlights a link between the unusual vortex core spectral signature and the pseudogap phase. 
       
\bibliography{VortexHTS.bib}

\onecolumn

\section*{Acknowledgments} 
We acknowledge discussions with S.~Sachdev, PM.~Bonetti, N.~Barišić, D.~Sunko, A.~Bussman-Holder, C.~Berthod and L.~Rademaker. We thank A.~Guipet and L.~Stark for their technical support.  
\paragraph*{Funding:} 
This research was supported by the Swiss National Science Foundation under grant \#200020\_182652.
\paragraph*{Author contributions:}
CR designed the experiment. TPS and IMA performed all the experiments and data analysis. GG grew the single crystals. CR, TPS and IMA discussed the results and wrote the manuscript. 
\paragraph*{Competing interests:}
There are no competing interests to declare.
\paragraph*{Data and materials availability:}
All the data discussed in the main text and supplementary materials are available on Yareta, the open data repository of the University of Geneva.

\subsection*{Supplementary materials}
\noindent Materials and Methods\\
Supplementary Text\\
Figures S1 to S7\\
References \textit{\cite{Wen2008JCG}}\\ 
\newpage

\begin{appendix}
     
\begin{center}
    \large \textbf{Supplementary Materials for} \\
    \vspace{0.2 cm}
    \large {
    \textbf{Vortex core spectroscopy links pseudogap and Lifshitz critical point in a cuprate superconductor} 
 } \\
 \vspace{0.2 cm}
\normalsize by Tejas Parasram Singar, Ivan Maggio-Aprile, Genda Gu, and Christoph Renner 
\end{center}

\vspace{0.2cm} 

\renewcommand{\thesection}{S\arabic{section}} 
\setcounter{figure}{0} 
\renewcommand{\thefigure}{S\arabic{figure}} 

\renewcommand{\thesubsection}{\arabic{subsection}}
\setcounter{section}{0}
\setcounter{subsection}{0}

\section*{Materials and Methods}

\section*{Materials}
The optimal doped single crystals ($T_{c}\approx91$ K) of Bi-2212 were grown using the floating-zone method \cite{Wen2008JCG}. Overdoped single crystals were obtained by annealing the optimal doped crystals in an oxygen gas environment (pressure $\sim$1000 -- 2000~bar) at 350~$^{\circ}$C -- 600~$^{\circ}$C for three days. The underdoped single crystals were obtained by annealing the optimal doped crystals in vacuum (pressure $\sim$ 0.001~torr) at 450~$^{\circ}$C -- 650~$^{\circ}$C for 100 hours. The superconducting transition temperature of different samples was characterized using the SQUID technique in presence of a 10~G field applied along the crystallographic c-axis.  

\section*{Experimental Details}
STM/STS measurements were performed at liquid helium temperature ($T \approx$ 4.4~K) in ultra-high vacuum (UHV) using a commercial SPECS Joule-Thomson Tyto instrument. The Bi-2212 single crystals were mechanically cleaved at low temperature ($T \approx$ 100~K)  in ultra-high vacuum ($P \lesssim$ 9 E-11~mbar) to obtain clean and atomically flat surfaces. The cleaving preferentially occurs between the  van der Waals bonded BiO planes, leading to the exposure of a clean BiO surface. STM measurements were performed using chemically etched, Ar$^{+}$ ion sputtered iridium (Ir) tips which were first calibrated on a Au(111) single crystal surface to obtain atomically sharp tips. The magnetic field was applied along the crystallographic c-axis of Bi-2212 using either an external coil (0 to 3 T) or a permanent magnet glued below Bi-2212 with a fixed field $B\approx$ 0.16~T. 

\section*{Supplementary Text}
\subsection{STM topography and tunneling spectroscopy as a function of doping $p$}
\begin{figure} 
    \centering
    \includegraphics[width=1\columnwidth] {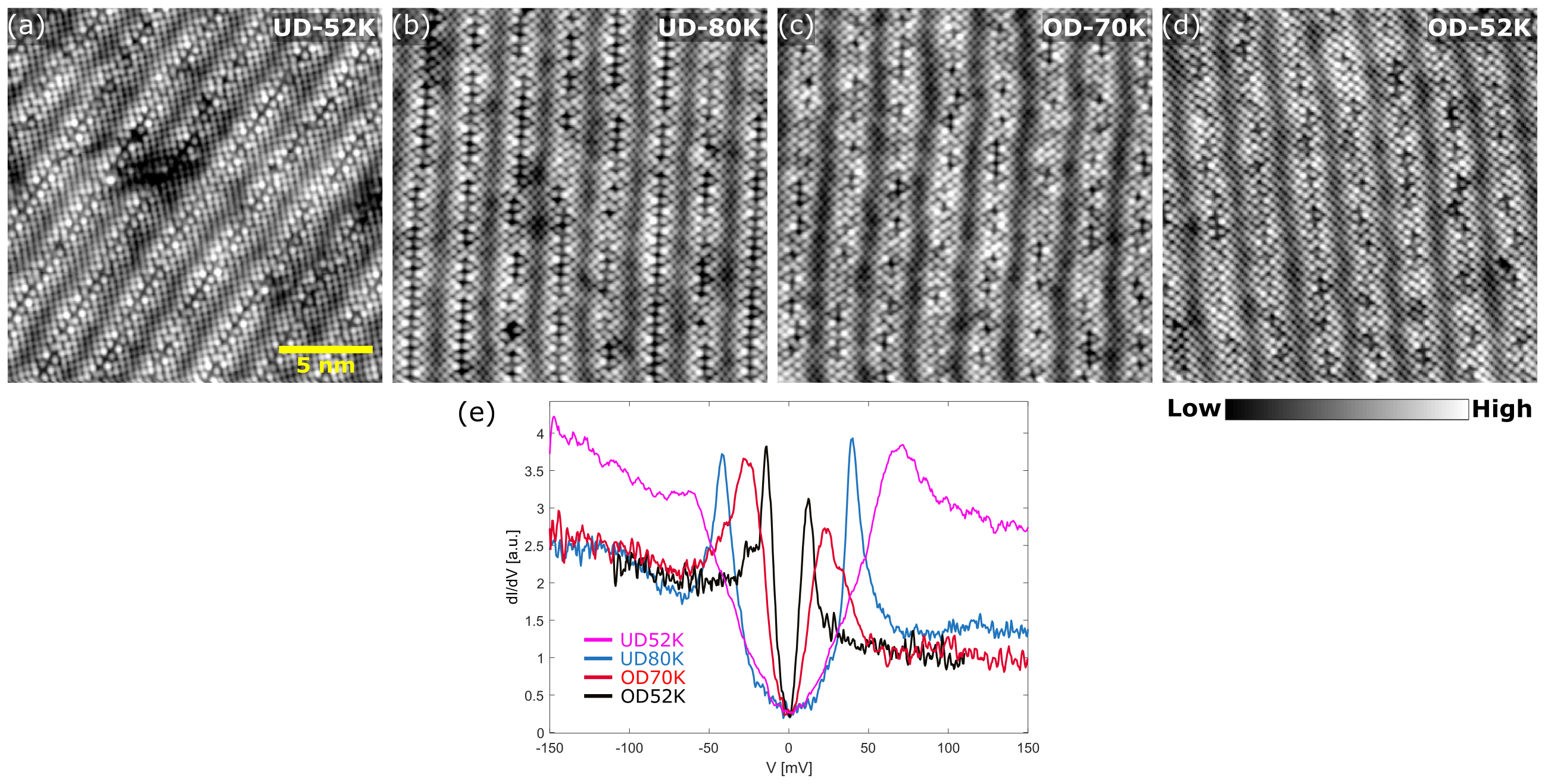}
    \caption{\textbf{STM topography and tunneling spectroscopy of Bi-2212 as a function of hole doping.} (a) Underdoped-52K. (b) Underdoped-80K. (c) Overdoped-70K. (d) Overdoped-52K. Scan conditions: $V_{\text{b}} =$ +250~mV, $I =$ 20~pA, $T\sim$ 4.4 K. (e) Representative tunneling spectrum for each doping level. The OD70K spectrum is normalized and vertically offset for clarity.      
}
    \label{fig: topovariation}
\end{figure}

With increasing hole content from heavily underdoped (UD52) to heavily overdoped (OD52), STM topographic images of the top BiO layer in fig.~\ref{fig: topovariation}a-d show that the superstructure crest evolves from a continuous serpentine-like feature to a more regular structure and sparse background impurities. STS shows prominent broad pseudogap features in underdoped crystals, which are replaced by sharp superconducting coherence peaks in heavily overdoped crystals (fig.~\ref{fig: topovariation}e). Finally, the spectra show a continuous evolution of the gap with increasing hole content, as expected from the Bi-2212 phase diagram. 

\begin{figure} 
    \centering
    \includegraphics[width=0.7\columnwidth] {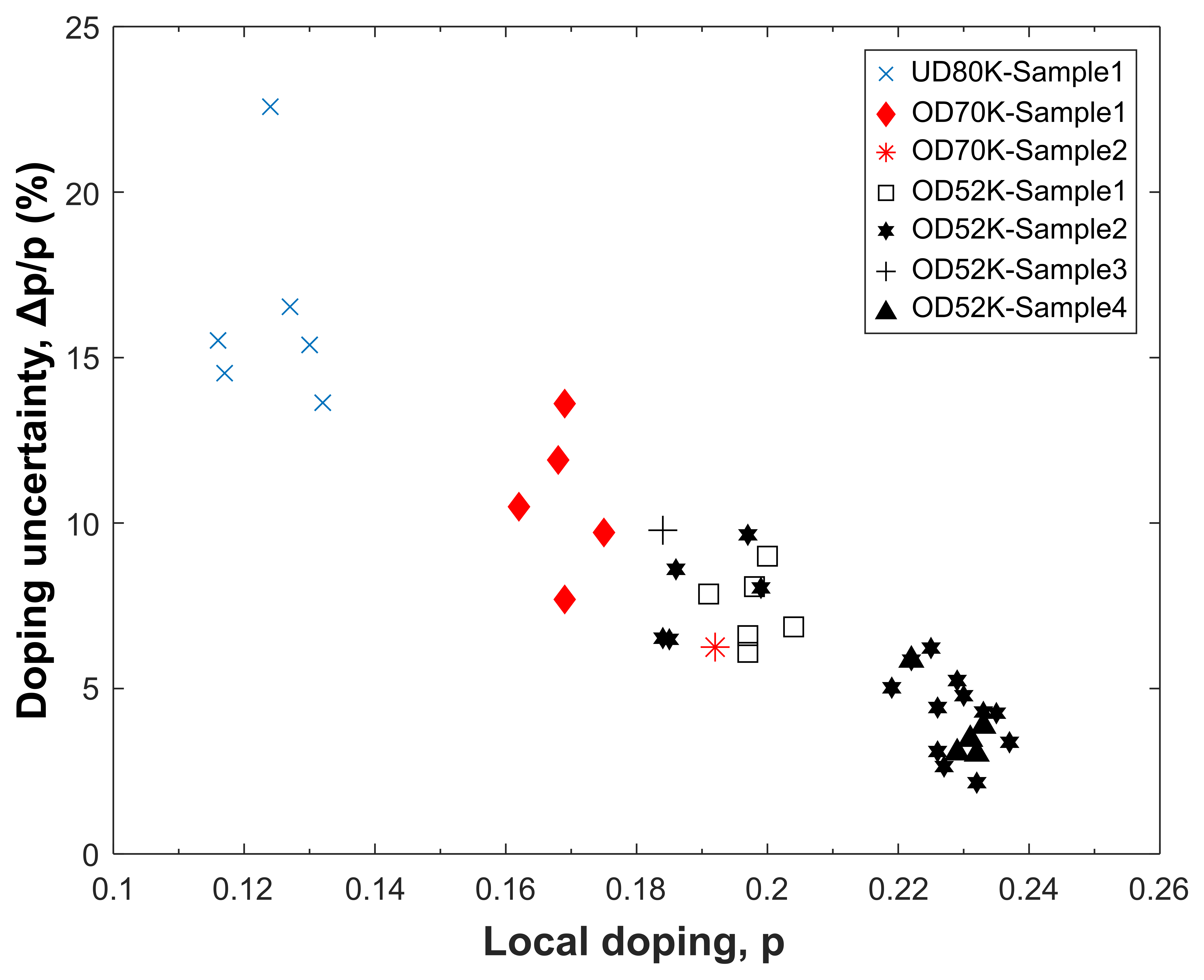}
    \caption{\textbf{Spread in local doping as a function of doping measured within a radius = 5.4~nm from the vortex center, but outside the core.} Different colors represent Bi-2212 batches with different bulk $T_{\text{c}}$ (UD80K, OD70K, and OD52K), whereas distinct symbols for a given color distinguish different samples from the same batch. All data points correspond to the measurements presented in the $H-p$ phase diagram in Fig.~\ref{fig: phasediagram} of the main text.   
}
    \label{fig: Doping_uncertainty}
\end{figure}

The relative gap variations in the vicinity of the vortex cores diminish with increasing hole doping, suggesting that Bi-2212 becomes overall more homogeneous in the overdoped regime (fig.~\ref{fig: Doping_uncertainty}).       

\begin{figure} 
    \centering
    \includegraphics[width=0.9\columnwidth] {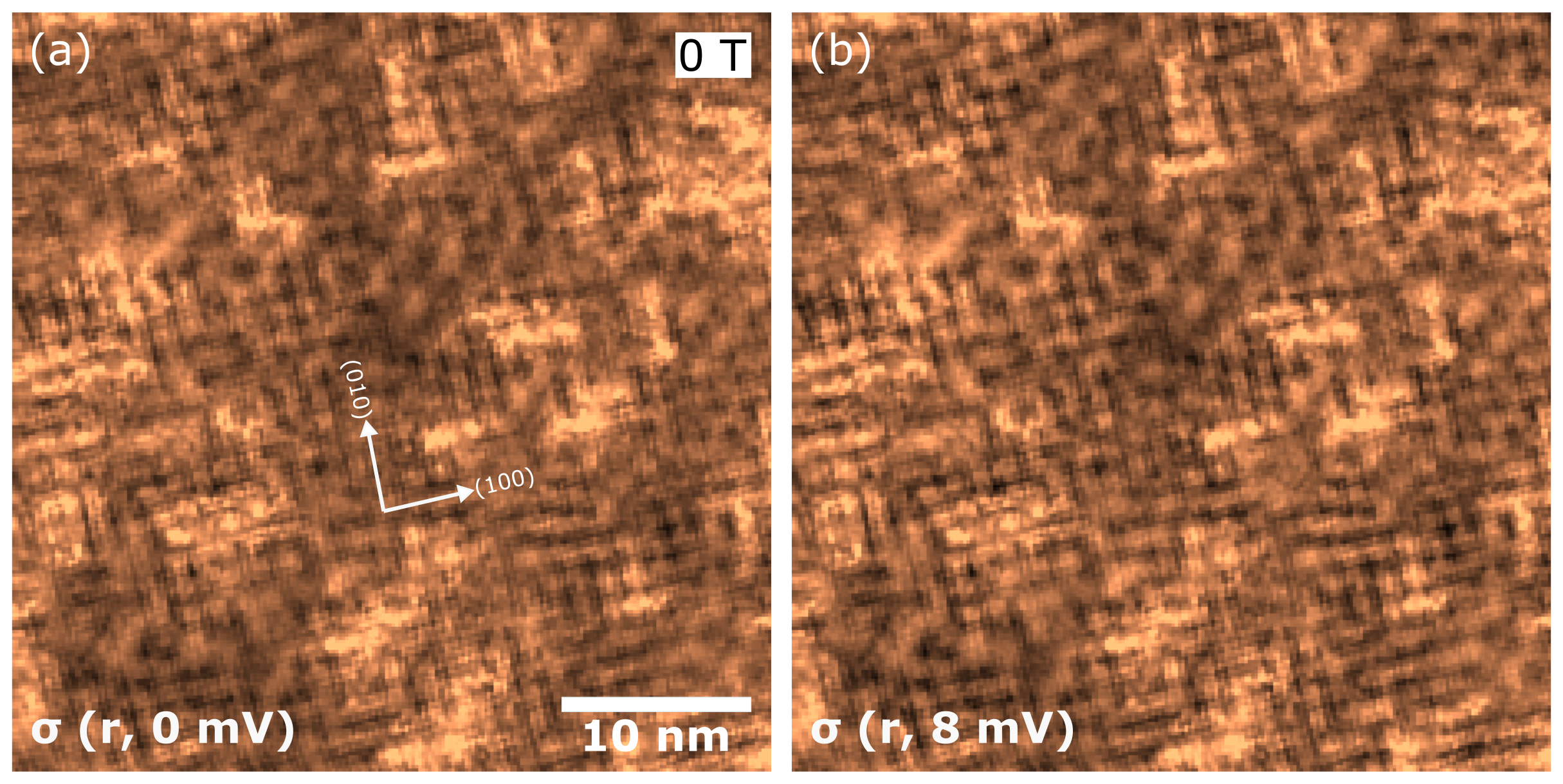}
    \caption{\textbf{Periodic conductance modulations in heavily underdoped Bi-2212 $T_c =$ 52~K.} (a) $dI/dV (r,V=0~mV)$ map. (b) $dI/dV (r,V=8~mV)$. The $\approx 4a_{o} \times 4a_{o}$ periodic modulations are present all the way to the Fermi energy, hampering the identification of a vortex core in these heavily underdoped regions. Scan conditions: $V_{\text{b}} =$ +100~mV, $I =$ 50~pA, $B =$ 0~T, $T =$ 4.4 K.         
}
    \label{fig: UD52KCB}
\end{figure}

\begin{figure} 
    \centering
    \includegraphics[width=0.8\columnwidth] {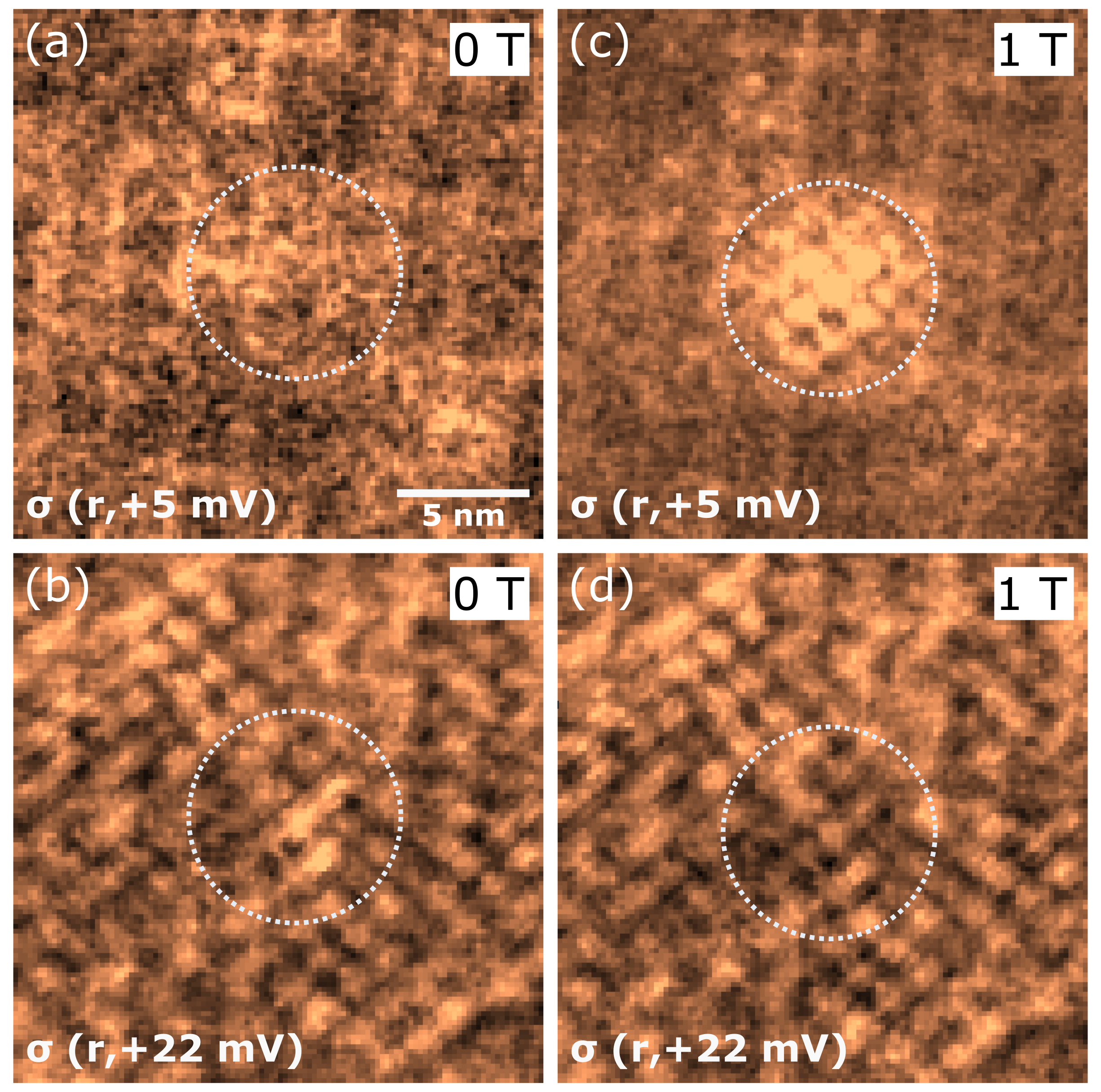}
    \caption{\textbf{Low and high energy conductance modulations in UD80 in the presence and absence of a vortex core (same data set as Fig.~\ref{fig: CBVTXUD80K} in the main text).} (a), (b) $dI/dV (r,V)$ maps at +5 mV and +22 mV at 0~T, respectively. (c), (d) $dI/dV (r,V)$ maps at +5 mV and +22 mV at 1~T, respectively. The white dotted circles indicate the vortex halo region aligned with atomic-scale accuracy. The slight offset between the 0~T and 1~T maps accounts for a small drift between the two separately acquired STS grid measurements. This data set clearly shows how periodic conductance modulations develop at low energy when a vortex core appears in a finite magnetic field, while there is essentially no difference at higher energy. Scan conditions: $V_{\text{b}} =$ +110~mV, $I =$ 50~pA.      
}
    \label{fig: UD80KVtx4suppl}
\end{figure}

It proved impossible to characterize vortices in heavily underdoped (UD52) crystals. This can be explained in part by the fact that periodic charge modulations persist to $E_F$ in the absence of a magnetic field (fig.~\ref{fig: UD52KCB}). Therefore, these regions appear very similar in presence or absence of a vortex core. Another possibility, which we cannot rule out, is that the vortices are mobile at $B<3$ T and escape from the STM tip while we are taking our measurements. At higher doping, the periodic charge modulations are suppressed at low energy (near $E_F$) when no magnetic field is applied. In this case, the vortex cores, whose halo hosts a periodic charge modulation up to $E_F$, become easily identifiable (Fig.~\ref{fig: CBVTXUD80K} and fig.~\ref{fig: UD80KVtx4suppl}).   

\begin{figure} 
    \centering
    \includegraphics[width=0.9\columnwidth] {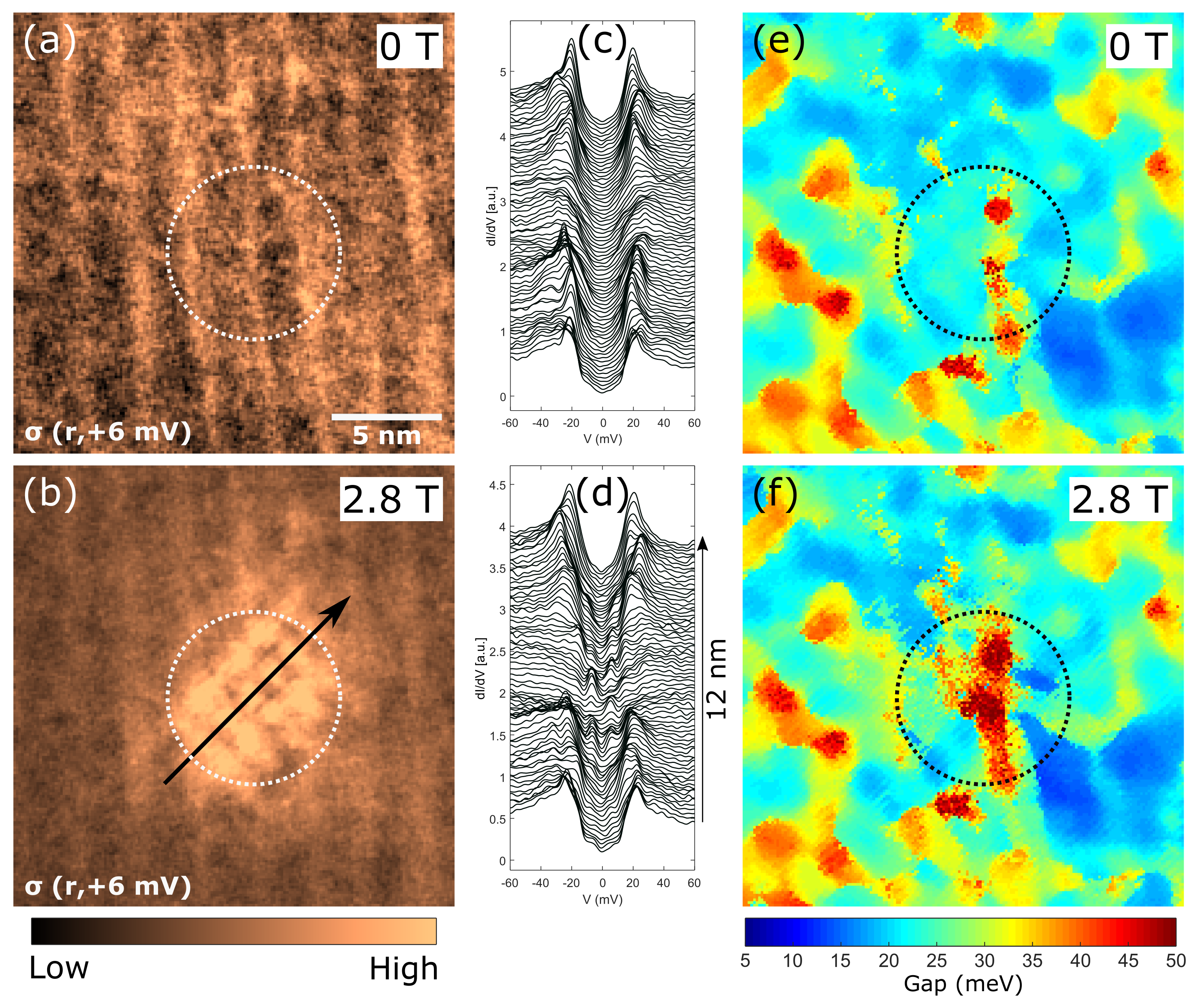}
    \caption{\textbf{Vortex core electronic structure in OD52 Bi-2212 at $p < p^{*}\approx$ 0.21 ($p_{\text{local}}= 0.20\pm$0.02) at 4.4~K.} (a) $dI/dV (r, +6~mV)$ map at 0~T. (b) $dI/dV (r, +6~mV)$ map in the same region in presence of a vortex at 2.8~T. (c), (d) STS conductance trace along 12 nm profile marked in panel (b) at 0~T and 2.8~T, respectively. (e), (f) Gap map in the same region as panel (a) and panel (b) at 0 T and 2.8~T, respectively. Here, the gap value corresponds to half the energy separation between two coherence peaks. Circles with radius 4~nm in panels (a), (b), (e), and (f) indicate the vortex halo region. Scan conditions: $V_{\text{b}} =$ +110~mV, $I =$ 50~pA. Note that the vortex investigated here is from the same OD52 crystal as the one discussed in Figure~\ref{fig: OD52KZBCPdecay} of the main text. 
}
    \label{fig: CBOD52vortex}
\end{figure}

The unconventional vortex cores are found at all locations on a Bi-2212 crystal where $p<p^*\approx0.21$, even in nominally highly overdoped crystals. One such vortex is shown in figure~\ref{fig: CBOD52vortex} at a site on OD52 where $p_{\text{local}}<0.21$. The electronic structure of the same region in presence ($B =$ 2.8~T) and absence ($B =$ 0~T) of the magnetic field shows clear appearance of the checkerboard pattern along with spatially bouncing sub-gap states. Furthermore, a gap map in the field of view demonstrates a local suppression of superconductivity over a coherence length distance from the core center (figs.~\ref{fig: CBOD52vortex}e, f). Note that the values within the vortex core do not represent real gaps since there are no coherence peaks to locate inside the core.  

\begin{figure} 
    \centering
    \includegraphics[width=1\columnwidth] {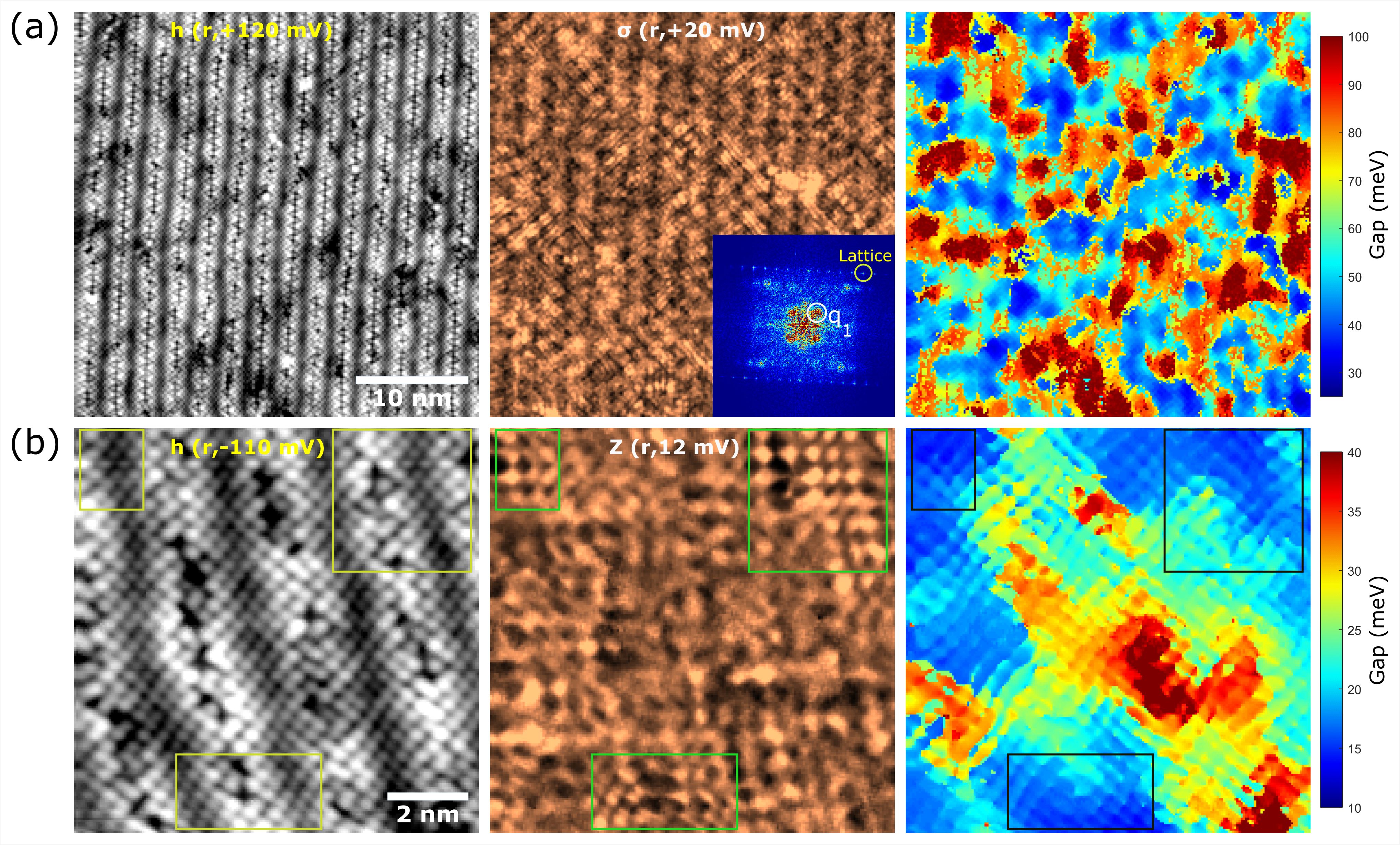}
    \caption{\textbf{Changing electronic modulations across $p^{*}\approx0.21$ in Bi-2212.} (a) Topography (left), conductance map $dI/dV (r,+20~mV)$ (middle, inset shows corresponding FFT), and gap map (right) acquired in the same field of view on a UD80 crystal. The $\approx 4a_{o} \times 4a_{o}$ checkerboard ($q_1$) modulation is clearly resolved in this sample where $p< p^{*} \approx0.21$ everywhere. Setpoint conditions: $V_{\text{b}} =$ +120~mV, $I =$ 50~pA, $B =$ 0~T. (b) Topography (left), $Z(r,12~mV)$ map (middle, where $Z(r,V) = dI/dV(r,+V)/dI/dV(r,-V)$), and gap map (right) acquired in the same field of view on a OD52 crystal. We clearly resolve strong $\sqrt{2}a_{o}\times\sqrt{2}a_{o}$ modulations in some regions outlined by rectangles. They are limited to regions where $p > p^{*}$ and where the gap is smallest in the gap map, as expected in highly overdoped locations. Setpoint conditions: $V_{\text{b}} =$ --110~mV, $I =$ 80~pA, $B =$ 0~T.}
    \label{fig: sqrt2modulation}
\end{figure}

The zero-field electronic modulations in Bi-2212 also evolve across $p^*$ (fig.~\ref{fig: sqrt2modulation}). Samples where the underlying Fermi-surface is open hole-like, conductance maps show strong $\approx4a_0\times4a_0$ checkerboard modulations --with precise periodicity depending on energy-- due to dominant anti-nodal scattering (fig.~\ref{fig: sqrt2modulation}a, fig.~\ref{fig: UD80KVtx4suppl}). However, at higher doping $p > p^*$, the checkerboard modulation is substituted by the $\sqrt{2}a_0 \times \sqrt{2}a_0$ pattern when the Fermi-surface becomes electron-like (fig.~\ref{fig: sqrt2modulation}b). The corresponding gap map at the same location shows that this pattern is restricted to the regions with the smallest superconducting gaps corresponding to locally highly doped regions.

\subsection{Making sure the ZBCP is an intrinsic vortex core feature} \label{ZBCPpushed}
\begin{figure}
    \centering
    \includegraphics[width=0.9\columnwidth] {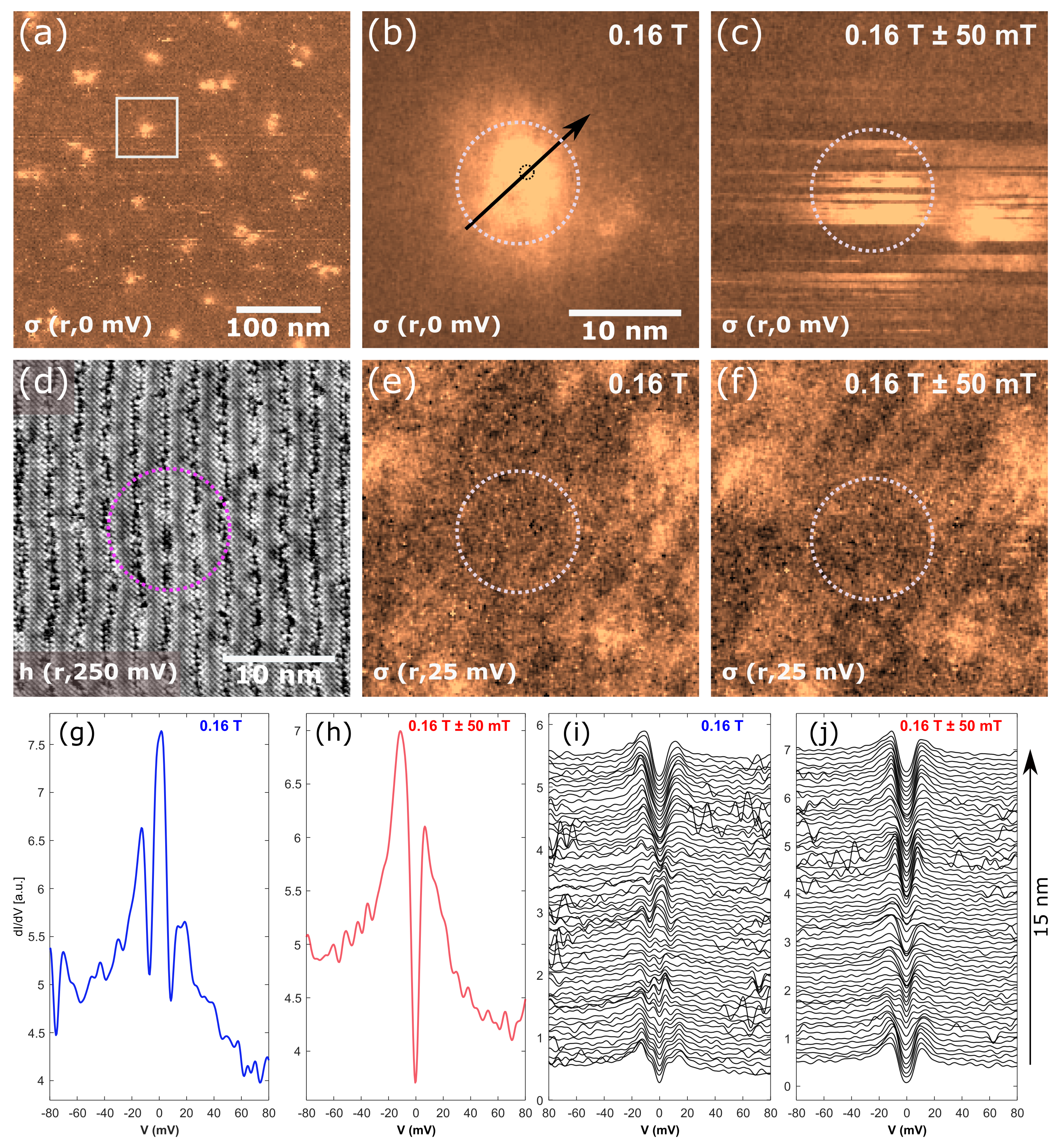}
    \caption{\textbf{Zero-bias conductance peak is an intrinsic vortex core signature in OD52 Bi-2212 at $p > p^{*}\approx$ 0.21.} (a) 400 $\times$ 400 nm$^{2}$ zero-bias conductance map displaying the vortex lattice at 0.16~T. (b), (e) High-resolution conductance map in a vortex region ($p_{\text{local}}\approx$ 0.23$\pm$0.01) at 0 mV and +25 mV, respectively. The position of this vortex in the lattice is indicated by the white square in panel (a). (c), (f) The same region as panels (b) and (e) probed after gently displacing the vortex by changing the applied field by about 50~mT. (d) Topography acquired at +250 mV in the vortex field of view. (g) STS spectrum at the vortex center displayed in panel (b). Spatial average performed over the area marked by black dotted circle in panel (b). (h) STS spectrum in the same area as that of panel (g), after displacing the vortex. (i), (j) Conductance trace along 15~nm line profile (marked in panel (b)) before and after moving the vortex, respectively. Circles in panels (b)-(f) indicate the position of the vortex halo region obtained from 0.16~T grid map measurement.     
}
    \label{fig: OD52KZBCP-vtxpush}
\end{figure}

To make sure the spectroscopic signature we find is due to a vortex and not some other feature such as a defect, we measure the exact same region, first with a vortex and then again after pushing the vortex away by slightly changing the applied magnetic field. The purpose is to make sure the spectral signature, and especially the ZBCP is an intrinsic vortex core signature. We first acquire a 400 $\times$ 400 (nm)$^{2}$ STS grid map showing several vortices (fig.~\ref{fig: OD52KZBCP-vtxpush}a). We then select one vortex (marked by the white square) to view it at two energies extracted from a high resolution STS grid map (figs.~\ref{fig: OD52KZBCP-vtxpush}b, e). The STM topography of the exact same region ($V_b =$ +250~mV) is shown in fig.~\ref{fig: OD52KZBCP-vtxpush}d. The tunneling spectrum in fig.~\ref{fig: OD52KZBCP-vtxpush}g is averaged over the dashed black circle in panel b and clearly shows the ZBCP at the vortex center. The full STS trace along the black arrow in panel b is shown in fig.~\ref{fig: OD52KZBCP-vtxpush}i. There is a clear conductance peak at zero bias at the vortex center, which splits in two peaks that shift to higher energy as a function of increasing distance from the vortex center.  

Next, we change the applied magnetic field by 50~mT, which shifts the vortex to the slightly brighter area in the lower right of the vortex in panel b. We view the same region again at two energies extracted from a high resolution STS grid map (figs.~\ref{fig: OD52KZBCP-vtxpush}c, f). While the map at +25~mV does not change much, which is expected since it is outside the superconducting gap, the map at 0~mV is modified. It is more unstable, suggesting that the vortex is initially shifting between the two bright positions seen in fig.~\ref{fig: OD52KZBCP-vtxpush}c. The tunneling spectrum in fig.~\ref{fig: OD52KZBCP-vtxpush}h is averaged over the same area as the one in fig.~\ref{fig: OD52KZBCP-vtxpush}g. Instead of a ZBCP, this location now shows a clear superconducting gap with sharp coherence peaks. The disappearance of the vortex is also clear from the full STS trace in fig.~\ref{fig: OD52KZBCP-vtxpush}j along the black arrow in panel b. Hence, the ZBCP and its splitting with increasing distance from the vortex core is a genuine spectroscopic feature of a vortex. On a side note, the trace in panel j shows that the vortex in panel b was sitting in a region where the superconducting gap is weaker, which is consistent with the system minimizing the energy cost of creating a vortex.

\end{appendix}

\end{document}